\documentclass[10pt, conference, compsocconf]{IEEEtran}
\usepackage{cite}

\ifCLASSINFOpdf
\else
\fi
\usepackage{url}

\usepackage{subfigure}
\usepackage{amssymb}
\usepackage{graphicx}
\usepackage{epstopdf}
\usepackage{fancybox, fancyvrb}
\usepackage{color}
\usepackage{amsmath}

\usepackage{graphicx}
\usepackage{mathptmx}

\usepackage{paralist}		
\usepackage{mdwlist}		
\usepackage{amssymb} 		
\usepackage{array}
\usepackage{parskip}
\usepackage{flushend}

\newcommand{\suppress}[1]{}  
\newcommand{\etal}{{\it et al}.}		
\newcommand{\eg}{{\it{e.g.}}}
\newcommand{\ie}{\emph{i.e.}}
\newcommand{\pw}[1]{{\fontfamily{cmss}\selectfont#1}}	
\newcommand{\dfp}{\pw{dispel4py}}

\newcommand{\sref}[1]{Sect.~\ref{#1}}				
\newcommand{\fref}[1]{Fig.~\ref{#1.fig}}				
\newcommand{\tref}[1]{Table~\ref{#1.tab}}				
\newcommand{\comment}[1]{}						
\newcommand{\shorten}{\vspace{-0.00cm}}

\definecolor{red}{rgb}{1,0,0}
\definecolor{maroon}{rgb}{0.5,0,0.2}
\definecolor{asparagus}{rgb}{0.5,0.5,0}
\definecolor{salmon}{rgb}{1,0.5,0.5}
\definecolor{blue}{rgb}{0,0,1}
\definecolor{blue(ncs)}{rgb}{0.0, 0.53, 0.74}

\newcommand{\hidempaNote}[1]{}						

\begin{document}
%
\title{VERCE delivers a productive e-Science environment for seismology research}



\author{\IEEEauthorblockN{Malcolm Atkinson\IEEEauthorrefmark{1},
Michele Carpen\'e\IEEEauthorrefmark{2},
Emanuele Casarotti\IEEEauthorrefmark{3}, 
Steffen Claus\IEEEauthorrefmark{4}, 
Rosa Filgueira\IEEEauthorrefmark{1},
Anton Frank\IEEEauthorrefmark{5}, \\
Michelle Galea\IEEEauthorrefmark{1},
Tom Garth\IEEEauthorrefmark{6},
Andr\'e Gem{\"u}nd\IEEEauthorrefmark{4}, 
Heiner Igel\IEEEauthorrefmark{7}, 
Iraklis Klampanos\IEEEauthorrefmark{1},
Amrey Krause\IEEEauthorrefmark{1}, \\
Lion Krischer\IEEEauthorrefmark{7}, 
Siew Hoon Leong\IEEEauthorrefmark{5}, 
Federica Magnoni\IEEEauthorrefmark{3},
Jonas Matser\IEEEauthorrefmark{8}, 
Alberto Michelini\IEEEauthorrefmark{3}, \\
Andreas Rietbrock\IEEEauthorrefmark{6}, 
Horst Schwichtenberg\IEEEauthorrefmark{4}, 
Alessandro Spinuso\IEEEauthorrefmark{8} and
Jean-Pierre Vilotte\IEEEauthorrefmark{9}}
\IEEEauthorblockA{\IEEEauthorrefmark{1}University of Edinburgh, UK. Email: Malcolm.Atkinson@ed.ac.uk}
\IEEEauthorblockA{\IEEEauthorrefmark{2}Supercomputing Applications and Innovation Department (CINECA), Italy}
\IEEEauthorblockA{\IEEEauthorrefmark{3}Istituto Nazionale di Geofisica e Vulcanologia (INGV), Italy}
\IEEEauthorblockA{\IEEEauthorrefmark{4}Fraunhofer Institute for Algorithms and Scientific Computing (SCAI), Germany}
\IEEEauthorblockA{\IEEEauthorrefmark{5}Leibniz Supercomputing Centre, Germany}
\IEEEauthorblockA{\IEEEauthorrefmark{6}University of Liverpool, UK}
\IEEEauthorblockA{\IEEEauthorrefmark{7}Ludwig-Maximilians-University, Germany}
\IEEEauthorblockA{\IEEEauthorrefmark{8}Royal Netherlands Meteorological Institute (KNMI), The Netherlands}
\IEEEauthorblockA{\IEEEauthorrefmark{9}Institut de Physique du Globe de Paris (IPGP), France}
}

%


\maketitle

\begin{abstract}
The VERCE project has pioneered an \mbox{e-Infrastructure} to support 
researchers using established simulation codes on high-performance computers 
in conjunction with multiple sources of observational data. 
This is accessed and organised via the VERCE science gateway 
that makes it convenient for seismologists to use these resources 
from any location via the Internet. Their data handling is made 
flexible and scalable by two Python libraries, \pw{ObsPy} and \pw{dispel4py}
and by data services delivered by ORFEUS and EUDAT.
Provenance driven 
tools enable rapid exploration of results and of the relationships 
between data, which accelerates understanding and method improvement.
These powerful facilities are integrated and draw on many other e-Infrastructures.
This paper presents the motivation for building such systems, 
it reviews how solid-Earth scientists can make significant research progress using them 
and explains the architecture and mechanisms that make their construction and operation achievable.
We conclude with a summary of the achievements to date and 
identify the crucial steps needed to extend the capabilities for 
seismologists, for solid-Earth scientists and for similar disciplines.

\end{abstract}

\shorten
\shorten

{\bf Note}: This is close to the final copy published by the IEEE in {\em The proceedings of the eleventh eScience Conference}, 2015. 
It contains minor corrections and additions, but 
please cite the conference paper\footnote{\url{http://conferences.computer.org/escience/2015/papers/9325a224.pdf}}.

%
\IEEEpeerreviewmaketitle

 \addtolength{\textfloatsep}{-5mm}
  \addtolength{\belowdisplayskip}{-5mm}
    \addtolength{\itemsep}{-5mm}

\shorten\shorten
\section{Introduction}\shorten

The EU VERCE project\footnote{EU VERCE, \url{http://www.verce-project.eu}, RI 283543.}, 
``\emph{Virtual Earthquake and seismology Research Community e-science environment in Europe}", 
has developed a comprehensive and integrated virtual research environment (VRE) for computational and data-intensive 
seismology balancing productivity gain with innovation potential. 
This has been pioneered with particular simulation models and data-driven seismology examples.
We report it here because the challenges it addresses are widespread and of growing prevalence and the 
solution strategy, which covers organisational and presentational as well as technical issues, is of wide relevance
and applicability. VERCE typifies current challenges in combining simulation results with observational
data as a research community grasps the opportunities presented by increased computational power and the growing wealth of data
while using existing resources and practices. Climate modelling has already addressed such issues under the
aegis of the IPCC. \sref{ResearchBehaviour.sec} presents comparisons and 
\sref{finale.sec} presents key lessons learnt and outstanding issues.


Seismology is one of the very few means of studying the sub-surface structure and phenomena of the Earth.
It provides an opportunity to model physical processes and compare the simulation results with surface observations 
from digital seismometers. The field is made more complex because it also addresses a sustained societal challenge 
of natural hazard mitigation \cite{PAGER2009}. We hypothesise that there are many cognate disciplines that 
will need to develop integrated e-Infrastructures supporting their communities in order to bring simulations 
representing their models and diverse observations into a conveniently used framework to enhance the 
productivity and capabilities of their research teams. We  anticipate that many of these will  have 
other drivers, such as applications to benefit society or business.

In such contexts it is essential to empower researchers with capabilities to explore options and to make rapid innovations.
Ideally they should be able to transfer the results from such creative contexts into large-scale production runs easily 
but that is constrained by two necessary compromises:
\begin{inparaenum}[\itshape a\upshape)]
\item	the production systems and major resources need to be well protected (against accidents and malevolence), and 
\item	some aspects of the work, such as preparing simulation codes to exploit the highest performance platforms well, 
	require intensive team effort involving different experts.
\end{inparaenum}
We report our path developing an effective VRE while handling these constraints in 	
today's rapidly evolving digital environment.

The present VERCE e-Infrastructure involves the following major elements moving from the researchers' point of 
contact to the contextual digital resources:
\shorten\shorten
\begin{itemize*}
\item	The \emph{scientific gateway} that is Web accessible from anywhere, provides an integrated view of all 
	available resources, handles continuity between sessions and supports collaboration with 
	shared data and methods, and with pervasive data access controls.
\item	The \emph{mapping to multiple Distributed Computing Infrastructures (DCIs)} handling identity management,
	authority controls, transformations between representations and protocols, and access to resources.
\item	The \emph{framework for computational science} that accommodates simulation codes, arranges for their use including 
	the supply of input data and the acquisition of results with associated metadata.
\item	The \emph{framework for data-intensive science} that delivers fine-grained composition of algorithms 
	and provides a smooth transition between development 
	and production with consistent semantics.
\item	An \emph{extensible coupling} with legacy and external services, and ingest of 
	non-standard or bulk data.
\item	A \emph{set of tools} to support validation, quality control and impact exploiting pervasive provenance records.
\end{itemize*}
\shorten\shorten

The paper has the following structure. \sref{ResearchBehaviour.sec} examines the requirements from researchers and their modes of using 
the VRE. It examines how this may evolve and extend into other disciplines. \sref{eInfrastructure.sec} outlines the architecture and key 
technologies used for each element of the VERCE VRE. \sref{experiencesBuildingTheVRE.sec} reports on the difficulties encountered setting 
up this coherent, European-wide accessible research platform---those overcome and those persisting.  
The benefits of building the VERCE e-Infrastructure and delivering its capabilities as a VRE are presented in \sref{VREbenefits.sec}.
\sref{finale.sec} summarises the progress made, evaluates it in a contemporary context and identifies priority research directions.

\shorten
\section{Use of the VERCE VRE} \label{ResearchBehaviour.sec}
\shorten\shorten
\hidempaNote{Cover: (a) seismometer networks and archives, (b) data preparation, (c) computational models, (d) using seismic data, 
(e) misfit analysis, (f) integration and application of seismology with other disciplines or non-research uses.
We promise above to ``examine the requirements from researchers and their modes of using 
the VRE and examine how this may evolve and extend into other disciplines."}

There are today large networks of digital seismometers, many of them permanently deployed, and some being moved periodically or deployed 
temporarily for a specific purpose. All of the seismic traces that they record are collected,	
standardised and made accessible under the aegis of the 
 FDSN\footnote{\url{www.fdsn.org}}, provided in Europe by ORFEUS and EIDA\footnote{\url{www.orfeus-eu.org/eida/eida.html}}. This 
 growing wealth of data is used to characterise earthquakes (\eg\  energy, location, depth and source mechanism). This serves 
 two main purposes:
 \begin{inparaenum}[\itshape a\upshape)]
 \item	alerting relevant responder authorities if the magnitude and proximity to populations warrants triggering further analyses 
 		immediately, \eg\ calculating tsunami scenarios for earthquakes in sea areas; and 
 \item	building catalogues of event properties, \eg{} GCMT\footnote{\url{http://www.globalcmt.org}} \cite{CMT-JGR81,CMT-PEPI2012} 
 		that are used by computational seismologists studying the Earth's structure and earthquakes' sources, and for estimating hazard and risk.
 \end{inparaenum}
As in many sciences, seismologists observe data, infer possible physical models at the origin of the data and then compare the results of the modelling with the observations.
They proceed  by formulating a ``\emph{mesh}" as a finite element model (FEM) of a 
volume to be studied---see \fref{FWPM2Inversion}.
This involves estimates of 
density and wave speeds for each element.
Building such meshes takes considerable effort and skill \cite{CaStLeKoPiTr08}. 
For each event in the region, the propagation of seismic waves is then simulated generating 
\emph{synthetic} seismograms for each seismometer. 
The observed and synthetic traces are then compared in a process called
\emph{misfit analysis}. The detected differences can then be back-propagated to refine the model in a process 
called \emph{inversion}. 
This can be done  
for all events in the region, and as the model of the Earth's structure improves it can be extended to finer resolution.
This is computationally intensive; current inversions require up to $10^{7}$ CPU hours \eg\
\cite{TaLiMaTr09,Fichtner09,Zhu:2012p216,Colli:2013jo}.
\begin{figure}[h]		
\centering
\shorten\shorten
\includegraphics[width=0.5\textwidth]{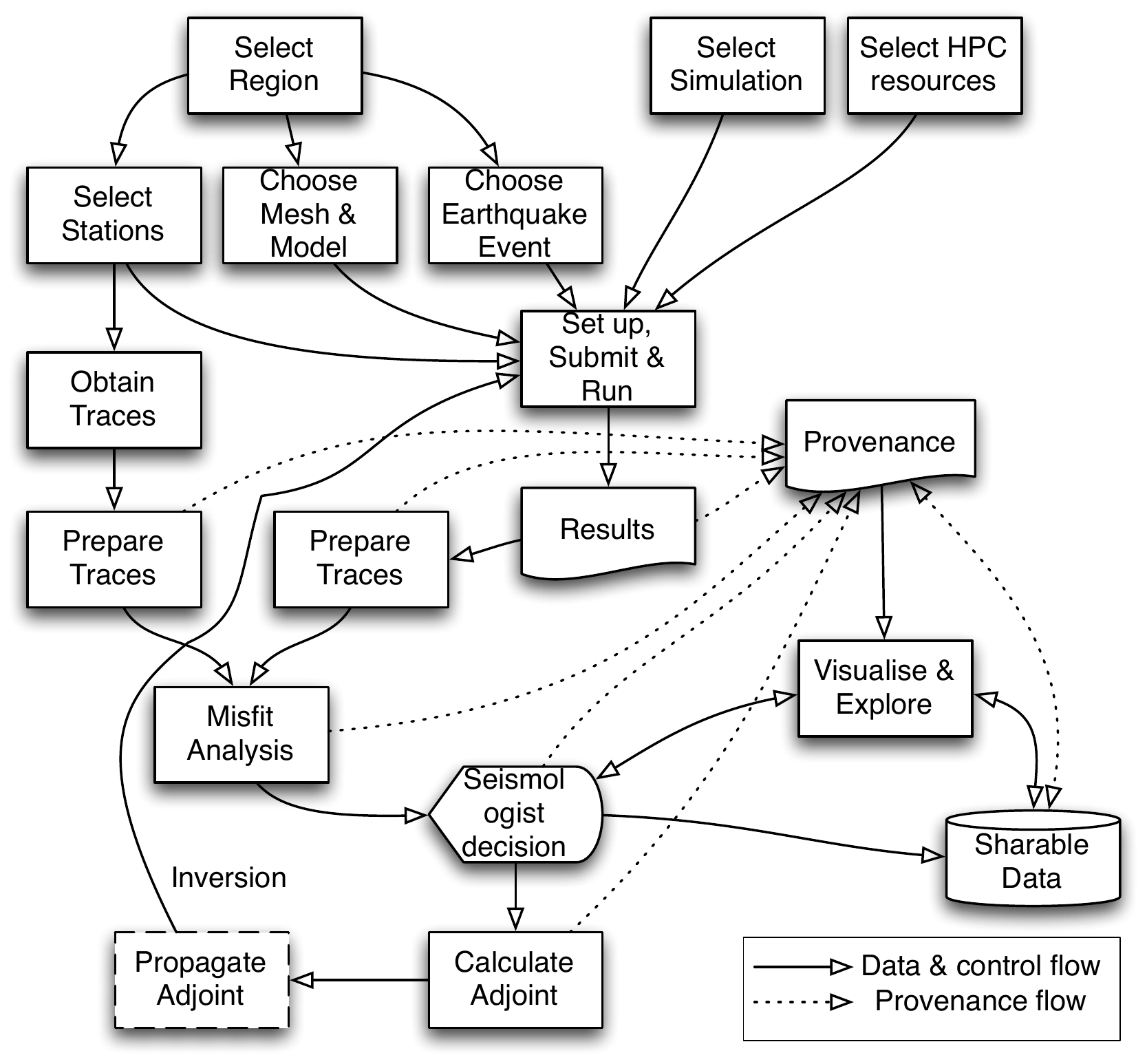}
\vspace{-1em}
\caption{Forward wave propagation, misfit analysis and inversion.\shorten\shorten}
\label{FWPM2Inversion.fig}
\shorten
\end{figure}

The forward wave propagation is well understood, but it requires considerable skill, effort and engineering to develop codes that simulate this efficiently 
by exploiting the power of modern HPC systems. A prime example, scalable to more than $10^5$ cores, is 
SPECFEM3D\_Cartesian\footnote{\url{geodynamics.org/cig/software/specfem3d}} \cite{Peteretal11},
which has been made much easier to use by the VRE. Prior to VERCE, each seismologist undertaking such a study needed to find their own resources,
understand the documentation of a simulation package, install and configure it, organise all of their runs, marshalling inputs and collecting results. 
The VRE automates all of this and presents it as a simple operation in the science gateway, where a seismologist
can select a region, choose an existing mesh or supply their own, to run the forward model of a chosen event for a specified time.
The science gateway enables seismologists to specify misfit analyses, selecting signal sources and comparison methods, as 
the next step in tomographic inversion.
The VERCE platform 
keeps provenance records to help scientists examine or repeat what they have done (\fref{FWPM2Inversion}).


Also, the collection and archiving of the continuous traces from seismometers has shown another route to seismic tomography. 
The previously disregarded ambient noise, which comes predominantly from ocean generated energy, can be correlated for 
long time intervals to yield the Green's function\footnote{The response at one seismometer to an impulse at the other.} 
between each pair of seismometers, \eg\ by stacking a year's worth of day-long trace correlations. 
This can reveal variation of the Earth's material properties with time. 
It is data-intensive, requiring $O({n^2})$ correlations of $n$ trace channels. 
VERCE provides a toolkit that automates and packages the required variations of this process and provides 
provenance tracking, saving seismologists many detailed data-management steps.


However, in both cases, scientists need to be able to select the region and frequencies of interest, and choose the processes to be run.
This particularly applies in both cases to the quality controls and data preparation on the recorded seismic waveforms \cite{Ringler+7-2015}
and to the post-processing, identification and preservation of result 
data together with any additional annotations that the seismologists require.
An extensive scientific Python library, \pw{ObsPy} \cite{ObsPy2011}, provides a repertoire of data-processing steps for this purpose and for ambient noise correlation, 
improving consistency and saving much re-implementation, thereby allowing seismologists to focus on their research rather than handle many details.
The \dfp\ Python library, described below, provides a data-intensive system to deliver scalability and flexibility for data-intensive processing \cite{FKAKS-Eescience2015dispel4py} 
with a simple abstract model.
These  empower researchers to formulate the data handing and transformations they require. 
Currently seismologists use these toolkits directly to explore their potential prior to packaged 
options being provided in the science gateway.


Seismic traces have further uses, \eg\ to detect landslides and glaciers calving, to monitor and analyse induced seismicity and other anthropogenic activity \cite{LACCM2014}, and to measure 
the effects of global warming on ocean wave energy. The set of solid-Earth sciences needs interworking of models and data, 
for example geodesic measurements, such as 
GPS data, LIDAR surveys, clinometers and photogrammetry, can be combined with seismic traces at fault-zone and volcanic observatories. 
They can then be compared with 
computational models of the geophysical processes. 
Similarly, models of mineral nanostructures, geochemistry and brittle-rock fracture 
can  be compared with laboratory induced fractures and geology in former fracture zones now exposed. 

An integrated VRE that incorporated such 
simulation systems and observational data will result in significant advances in geosciences. 
The integration of geophysical resources planned by EPOS (the ESFRI infrastructure for solid-Earth sciences\footnote{\url{www.epos-eu.org}}) 
will provide the critical organisational context and many of the key resources. The geo-specific  
examples portend a much wider application of the VERCE approach that will require access to observational and experimental data. 
It will depend on the simulation codes and data analyses being made efficiently scalable, and on provision of resources for simulations and 
for data. 

This will depend on a well-founded incorporation of diversity in software, data, computational systems and institutional 
goals---an organisational as well as technical challenge to deliver a coherent research environment in the context of loosely coupled systems and 
independent institutions that serve many other goals. Research changes understanding and therefore goals.
Commercial pressures and innovations drive changes in technologies.
Diversity will therefore continue. It will not be subdued by emergent business models or standards; 
these become extra members of the digital ecosystem in
which research must thrive.
Without an integrating framework that embraces inherent and evolving diversities, researchers will 
continue long and tedious struggles facing diversity alone, never reaching the full power available. 
VERCE overcame organisational and technical impediments to reach critical mass, diversity, scope and quality.
VERCE pioneers a shared investment delivering a coherent research environment in such an evolving diverse ecosystem and  
makes the advantages of the coherence and the technical strategy available to others. 
For example, to assess and respond to natural hazards  
or to respond to the societal challenges for renewable energy and resources.

The SCEC project\footnote{\url{scec.usc.edu/scecpedia/Geoinformatics_Project}} strives for
better prediction of strong ground motion, which requires better 3D models. It
includes work on simulation codes, whereas, in principle,
VERCE hosts independently developed code. It differs from
VERCE in delivering the
services from one HPC platform and site, thereby simplifying
the e-Infrastructure required, but missing the potential
combination of multiply owned resources. Its users may
have less opportunity to provide their own finite-element
models and to shape their own collection, preparation and
post-processing of seismic observed and simulated data---a
tradeoff still being explored between flexibility, complexity and community ownership
in VERCE.

The Shake Movie service run by Princeton University\footnote{\url{global.shakemovie.princeton.edu}} uses SPECFEM\_GLOBE
to generate simulations corresponding to recent earthquakes. Here users can access the simulation results but cannot set up and steer 
their own simulations.


\shorten
\section{The VERCE e-Infrastructure} \label{eInfrastructure.sec}
\shorten\shorten
\hidempaNote{Cover the 6 elements given above as necessary to support the researcher behaviours identified in \sref{ResearchBehaviour.sec}.
We promise above to ``outline the architecture and key 
technologies used for each element of the VERCE VRE."}

\fref{overviewArchitecture} shows the major components of the e-Infrastructure needed to deliver the VERCE VRE. 
Simply integrating, presenting and tailoring technologies does not complete a VRE.  That requires communities driving the research, 
collections of relevant data, application-software tuned and maintained to  
meet the latest research requirements and to exploit hardware advances, and teams of 
ICT experts maintaining its leading capabilities. However that framework and its usability and sustainability is key to attracting these 
researchers, enabling their collaboration, creating effective interplay between ICT experts and domain-focused researchers, gaining access to the resources required
and amortising costs over sufficiently broad communities. 
As already noted this has to embrace challenging diversity in virtually every element.

\begin{figure}[h]		
\centering
\shorten\shorten
\includegraphics[width=0.5\textwidth]{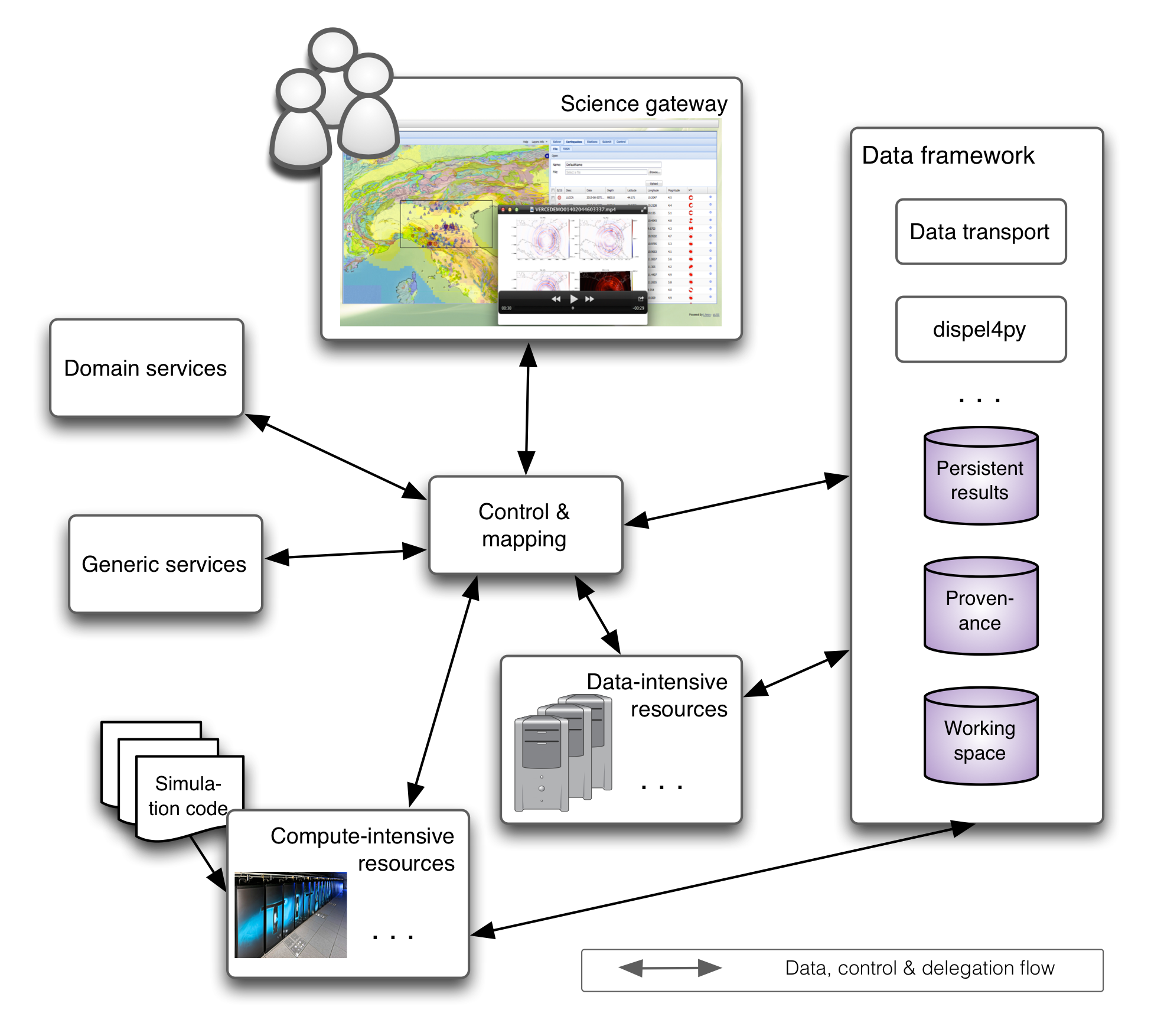}
\vspace{-1em}
\caption{Overview of the VERCE e-Infrastructure architecture.\shorten\shorten}
\label{overviewArchitecture.fig}
\shorten\shorten
\end{figure}

\shorten
\subsection{The VERCE science gateway} \label{VERCEgateway.ssec}\shorten
\shorten
\hidempaNote{The structure, facilities and implementation of the science gateway}

A science gateway has to support all the required modes of interaction with all aspects of the VRE needed by the geographically dispersed and often mobile 
research community. It has to support training and induction into this community, organisation of sharing, distribution of responsibilities and attribution of credit.
It needs to provide convenient and understandable access to tools and functions that enable all stages of the research methods, such as those illustrated above, 
from their creation to their application to build evidence for conclusions, publications and actions. Whilst enabling the domain scientists to be as productive and 
agile as possible is the highest priority, the work of the technical teams supporting them should also be facilitated, with many of the technical chores 
automated to improve efficiency, accelerate delivery of new facilities and reduce error rates.

The science gateway is built on \pw{SCI-BUS} \cite{SCI-BUSbook2014} which itself builds on Liferay\footnote{\url{www.liferay.com}} 
and \pw{gUSE/WS-PGRADE} \cite{Kacsuk2012}, which provide 
an Internet portal framework and a task-oriented workflow system respectively. 
Their wide use promises shared future maintenance and their maturity and integration implied reliability.
The gateway organises and structures the set of available options as portlets with 
drop-down menus of the required functions, different sets of functions being visible depending on the authority and mode of the logged-in user. A typical function will 
be illustrated using the modelling of seismic wave propagation introduced above---see \fref{FWPM2Inversion}.  Via standard geospatial interfaces  seismologists choose
their region of interest, which then exposes the existing models and available seismic stations,
from which they choose a mesh and velocity model.
They then	select a set of seismic stations to be used for simulations and misfit analysis, 
and select a relevant event, \ie\ an earthquake record from GCMT, 
	or from other seismic source solution catalogues, e.g. TDMT\footnote{\url{cnt.rm.ingv.it/tdmt.html}} from INGV \cite{ScTiMi09}.
Users can supply their own mesh, velocity model, seismic observations and pre- or post-processing steps at each stage.
They then submit a  simulation code run (currently using SPECFEM3D).
The gateway 
\begin{inparaenum}[\itshape a\upshape)]
\item	supports region selection by geographic name, by map zoom and by return to previous session contexts, 
\item	it provides structural-model and mesh upload or selection from an existing library of meshes and models,
\item	seismic stations are selected by clicking their icon on the region map,
\item	the gateway queries the GCMT or TDMT Web services and displays a list of seismic events.
\end{inparaenum}

To perform the submission, it uses the \pw{SCI-BUS} \pw{DCI-Bridge} \cite{DCI-BridgeChapter2014}, which handles many aspects of submitting a workflow to different
targets, transporting data and security credentials, manages job submissions and collects status updates---an example of reducing the ICT chores.
Arrangements to match the simulation code, the workflow, the credentials and the data handling, still require substantial effort from ICT experts in current target systems---see below.
Obtaining relevant traces and station metadata for the identified stations for the propagation time of the waves for each event is implemented as queries to seismic archives.
The preparation (\emph{pre-processing}) of these selected traces and the synthetic traces then uses a common \dfp\ pipeline that seismologists have previously designed and installed.
This can also be fed to a misfit analysis, also using \dfp, that yields visualised results. A seismologist makes a decision as to whether these are worth preserving and whether to 
use the computed misfit measurements to validate the geological model or as a source of 
back-propagated wavefield that interacting with the forward wavefield, highlights the required corrections to improve the structural model.
This back-propogation step 
is not currently set up in the VERCE gateway. 

To accomplish the above functionality the gateway has to support all of the user interactions by generating graphical presentations and by interpreting users' actions.
The interpretation involves the generation of workflows shaped by those actions and by the session history. It maps 
from workflow templates (developed by experts) to target-specific workflows 
and job descriptions in submission format. 

The sites all require that identity, security and authorisation issues are handled  
and their diversity must be addressed. 
The gateway attempts to provide uniform access to 
a federation of sites.  
Two  strategies are available:  integration of user identity and account data into a
centralised database or use of a decentralised schema. For example,
PRACE\footnote{\url{www.praceri.eu}} distributes 
user-account information across \pw{LDAP} servers replicating
users' information on each site, which  
can then have 
local policies. This forces administrators to maintain their own
authentication service.
Consequently, to use 
resources  users have to register with each site via its system
administrators. 
General software products support the centralised approach; \eg\  Shibboleth\footnote{\url{shibboleth.net}} 
seeks a complete solution to authentication using \pw{SAML} \cite{SAMLV2-2008}
to allow people to sign in with just one identity to independently run systems on which they have an account.
An identity provider supplies information as \pw{SAML} 
assertions, while resource providers use this information to adjudicate access. 
This offers flexibility but sufficient trust and wide adoption is needed.
For VERCE such consistency is not yet available, \eg\ 
\pw{GridFTP} (data transport)  uses  proxy certificates that do not support \pw{SAML}.
UNICORE\footnote{\url{www.unicore.eu}} (middleware used by some HPC providers) supports Explicit 
Trusted Delegation (ETD) \cite{DBLP:journals/fgcs/BenedyczakBBMS11} while
iRODS\footnote{\url{wiki.irods.org}} (used for federated data and metadata storage) provides authentication via 
Pluggable Authentication Modules (PAM) \cite{PAM2003}.

Addressing such diversity is essential. The computations require substantial resources and different researchers have 
different sites at which they have allocations.  These change as a result of competitions, policy changes and transfer between projects. 
Careful tuning of each simulation code to exploit the latest 
advances in parallel architectures is needed; consequently for each run different codes have different targets where they perform best. 
Economic and technical motivations warrant service amortisation over multiple communities, \eg\ data transport and
preservation delivered by 
EUDAT\footnote{\url{www.eudat.eu}}. We revisit such issues in \sref{experiencesBuildingTheVRE.sec} and \sref{finale.sec}.

\shorten\shorten
\subsection{Composing DCIs} \label{multiDCIs.ssec}
\shorten\shorten
\hidempaNote{Why are multiple DCIs needed? What are the challenges? How are these met?}

We use diverse HPC e-Infrastructures and access remote storage systems such as \pw{B2SAFE}\footnote{\url{www.eudat.eu/services/b2safe}} from EUDAT and
FDSN services from EIDA; these use different DCIs. Further diversity is required  
\begin{inparaenum}[\itshape a\upshape)]
\item	to transition between phases of research from exploration and development, on small convenient facilities, to production runs, and 
\item to efficiently handle loads across the compute-intensive to data-intensive range.
\end{inparaenum}
Changes in the user expectations  appear as a consequence, 
\eg\ they expect a simple and familiar log-in to their own machine or institution's DCIs.
During field work and when collecting observations from remote temporary deployments they may work 
disconnected from the Internet and independently from reference services.
Normally they expect a closely coupled system with automated access to many services.
Transitions between such contexts require bulk ingest of new observations into the framework.
Simulation code optimised to fully exploit advanced parallel architectures needs to be located on the architecture for which it was optimised for production 
but can be hosted on lower performance clusters and elastic clouds for smaller-scale development runs. 
Data-intensive workloads (see \sref{DataIntensiveFramework.ssec}) will under-utilise CPU resources on HPC platforms, so data is moved to appropriate clusters for processing.
Specialised data-intensive clusters, such as the Terracorrelator\footnote{\url{www.wiki.ed.ac.uk/display/Terra/Terra-correlator+wiki}}, 
outperform standard clusters and clouds \cite{dispel4py+IJHPCA-2015}. Parallel I/O and dynamic data compression is needed 
for production scale runs \cite{986} but standard data operations found locally are adequate for test runs.
Commercial clouds are used for initial deployments, \eg\ OPENSHIFT for the GCMT FDSN service development.
VERCE has to integrate diverse DCIs---DCI-Bridge \cite{DCI-BridgeChapter2014} substantially reduces 
the effort needed.

\shorten\shorten
\subsection{The computational science framework} \label{CompSciFramework.ssec}
\shorten\shorten
\hidempaNote{What is needed to run simulations? How does it vary with DCI and codes? How is this set up to (a) be easy to use, and (b) be easy 
to accommodate new simulation codes and models?}

Researchers need to use institutional and national facilities where they have routine access, and transfer to the highest performance facilities, such 
as the PRACE sites, when they are granted time to tackle demanding simulations. They require help from the VERCE framework 
moving their work, with its associated code sets, workflows and data, between computational facilities, and they need to retrieve results and 
provenance records into their sustained working context. Beyond facilitating the transfer of work and data, this requires 
careful tuning of code and workflows for the highest end context. Once systems and software engineers with specialist knowledge have done such preparations   
the VERCE framework supports the transitions and presents the computational services 
uniformly to researchers entitled to use them. This entitlement is facilitated for training and induction, but it remains challenging for 
significant use because:
\begin{inparaenum}[\itshape a\upshape)]
\item	the cost, particularly in energy, of such use is very substantial, 
\item	the risk of malevolent misuse of such powerful systems has to be minimised, and
\item	constraints are imposed on owners of such resources.
\end{inparaenum}

\shorten\shorten
\subsection{The data-intensive framework} \label{DataIntensiveFramework.ssec}
\shorten\shorten
\hidempaNote{The \pw{dispel4py} motivation, capabilities and mappings}

The data-intensive facilities build on the \dfp\ framework and on a set of common data services---see \fref{overviewArchitecture}. 
\dfp\ enables scientists to develop their own data-intensive 
applications using Python, familiar tools and their normal work environment, and to map automatically to a range of target systems adapting dynamically to the 
scale of data \cite{FKAKS-Eescience2015dispel4py}. There is a core library to support initial development and mappings to four scalable target enactment models:
\pw{MPI}~\cite{anon94mpi}, \pw{Apache Storm} clusters\footnote{\url{storm.incubator.apache.org}} and shared-memory multi-core machines 
using \pw{Multiprocessing} (a Python package for concurrent interpretation) and \pw{Apache Spark}\footnote{\url{spark.apache.org}}.     	
These mappings employ established middleware to benefit from its substantial development effort and implementations on multiple DCIs. 
\dfp\ integrates with provenance mechanisms---see \sref{ProvenanceTools.ssec} and has a \pw{registry} to promote method and component description and sharing.

\hidempaNote{Describe concepts of \dfp\, illustrate mappings, reference performance measures and pr\'ecis registry.}
\dfp\ encodes data-intensive methods as a graph of \emph{operations} coupled together by \emph{data streams}. 
The operations are instances of \pw{Processing Elements (PEs)} that consume units of data from their inputs and emit units of data on their outputs.
They may have zero or more inputs and zero or more outputs. The units of data are application dependent, and may be of any size, from a small scalar to a 
complex object, such as a multi-dimensional array with its metadata.
The data streams may use any communication mechanism. They may use compression and provide sufficient buffering.  
They deliver, preserving order, the units of data 
supplied by an output or external source to all of the inputs to which they are connected.
The operations are instances of Python classes and the stream connections are made by Python operations on these classes.
Initial sets of these \pw{PE} classes can be downloaded as a Python library or obtained from the \pw{registry}. 
Both scientists and experts extend the set of \pw{PEs} by writing their own algorithms or wrapping legacy algorithms using Python's extensive scientific libraries and 
call outs to other languages. A \pw{PE} may also wrap a subgraph so that it may be used as an operation.
The connections use local in-memory communication if this is available, and direct communication over local or wide-area protocols; 	
they do not go via disk I/O unless buffering overflows. Explicit writes to disk for final results or diagnostics require parallel I/O in production contexts.	

When the script is run, the graph is built as Python objects. In a development context this is interpreted locally. When larger scale runs are required, the 
graph is optimised, \eg\ \pw{PEs} are clustered to minimise data transport within loading constraints, the target DCI and intermediate technology is chosen, 
and the graph is mapped onto these targets with appropriate pipelining and parallelisation.
The use of data streaming has three benefits:
\begin{inparaenum}[\itshape a\upshape)]
\item	the low overheads, particularly for local coupling, means that it is efficient for scientists to compose low-cost steps, called ``{\em fine-grained workflows}", 
\item	the direct handling of streams means that scientists can develop methods for live continuously flowing observations, and
\item	the load placed on the increasingly limiting bottleneck of disk I/O, Kryder's law \cite{KrydersLawSciAm2005}, is minimised.
\end{inparaenum}
The abstract form of \dfp\ together with automated mapping means that scientists are not distracted by technical details and 
their encoded methods are not locked into a particular target. The automated mapping also relieves experts of the tasks of
optimising and mapping for particular DCI architectures. 
The choice of DCI target and mapping is critical for production performance, \eg\ Filguera \etal\ report for a 1000 station cross correlation, 
the Terracorrelator running slightly faster than SuperMUC\footnote{\url{www.lrz.de/services/compute/supermuc/systemdescription}} 
using \pw{MPI} and 5.5 times faster than Amazon EC2 \cite{dispel4py+IJHPCA-2015}. This becomes more significant when we note the 
low cost and energy consumption of the Terracorrelator---many organisations are now acquiring such machines designed for 
data-intensive loads \cite{Givelberg:2011:ADC:2110217.2110226}. The widespread availability of Python, including on super-computers,
increases the number of potential targets and reduces requirements for ICT experts to perform installation.


The  \pw{registry} provides consistency via a Web-based registration service for \dfp\ components. 
It encourages collaboration amongst researchers by providing a context where they can search, download, revise, annotate, register and reference \dfp\ constructs. 
It uses standard representations to enable interaction with other workflow repositories, \eg\ \pw{Wf4Ever} \cite{wf4everObjectModel2014}.
The  \pw{registry} stores \pw{PEs}, \pw{functions}, \pw{connections} and workflow graphs. 
Using components has been made transparent by overriding the Python  \pw{import} keyword so that \dfp\ fetches non-local items from the \pw{registry}. 
The \pw{registry} provides  workspaces to allow researchers to define components and execute experiments controlling their visibility. 
A workspace represents a snapshot of the library and other, \eg\  domain-specific components.
Workspaces are organised as a hierarchy, with new components and links to parent workspaces.

VERCE also needs to integrate diverse data-management systems, including file systems on the many computers in use, DBMS and shared data services.
These include multiple data storage systems and services for finding, referencing and moving data between them, as shown in \fref{overviewArchitecture}.
\hidempaNote{continue here with an outline of the data facilities.}
Data needs to be in the right place at the right time, \eg\ on a cluster about to run a simulation if it is input and removed from that cluster 
immediately the job completes if it is a result set. 
This requires careful scheduling of many data movement operations; these are automatically incorporated into workflows, saving scientists 
from such details. The underlying mapping system needs to choose the right mechanism for the job, \eg\ \pw{GridFTP} for large volumes as it 
handles recovery and retry, but \pw{ftp} and \pw{http} for small transfers. In the longer term, shared services provided by EUDAT and 
generic technology such as \pw{Data Avenue} \cite{DataAvenueSCIbusBookChap5-2014} should reduce the ICT chores involved, but 
understanding and specifying the required movements, resource management, authorisation, preservation and persistent identification will still be needed.
The distributed system for results and their metadata is built on iRODS \cite{DBLP:conf/policy/ConwayMRN11}, and this links closely with 
provenance, described below, which uses MongoDB\footnote{\url{mongodb.org}} as its distributed NoSQL DBMS. \nocite{DBLP:conf/policy/2011}
The broad user bases of iRODS and MongoDB motivated their selection, support for iRODS from EUDAT helped sustainability, and MongoDB 
combined flexible data strutures with scalability.

\shorten\shorten
\subsection{Connecting external facilities} \label{ExternalFacilities.ssec}
\shorten\shorten
\hidempaNote{No subdiscipline is an island. How do we make it easy (a) to bring in and use information and methods from other contexts, and
(b) enable other disciplines and groups use VERCE information and methods?}

These fall into two groups:
\begin{inparaenum}[\itshape a\upshape)]
\item	domain-specific services, in this case seismology, geoscience and geographic services, and
\item	generic services that support these and other communities, and facilitate interworking between software systems.
\end{inparaenum}
These categories form a continuum as some services, such as those providing map and political data are widely used.
The challenge is to achieve a balance between making it easy for specialists to use their particular sources of data and
facilitating interdisciplinary boundary crossing---both access to other disciplines' data from VERCE and access
to VERCE's data from other disciplines.
\hidempaNote{Alessandro plans to provide some material here.
Aless: Section III - E. Some general thinking on practices supporting cross disciplinary awareness. Maybe some more details on domain-specific services is needed? Do we have to be technical?} 

The immediate needs of the research activities described above are mainly concerned with getting access to seismic observations, 
geological models, the meshes and earthquake event parameters. The seismic traces and the metadata associated with the seismic stations, 
often come from FDSN compliant services,
however they can come from {\it ad hoc} deployments organised for specific purposes, \eg\ to study structure exploiting aftershocks, or as 
international exchanges of large collections of data, \eg\ the Japanese data relating to the 2011 T\-ohoku earthquake. The FDSN sources can 
be interrogated by Web services, but the other data needs careful ingest into the system, to adopt relevant standards, to generate  
necessary metadata and to build indexes that allow their use via the same framework and underlying workflows.
This is an open-ended requirement that has to involve experts who understand the form and origin of the data. However, they are 
provided with a set of processing elements that they can select, combine and parameterise and if necessary augment using \dfp\ to establish 
such \emph{data ingest} methods, which can then be reused and refined. These tools are used directly at present, while understanding about
how best to package them in  the science gateway develops.

The geological models and meshes are less standardised and strongly depend on the simulation code adopted. 
The mesh generation by a computational seismologist requires topography and geological data, which is often obtained from external services, FEM building tools, 
\eg\ CUBIT\footnote{\url{cubit.sandia.gov}}, and knowledge about how mesh properties affect running times and model resolution. 
It is difficult to integrate a meshing generation toolkit into the VRE due to licensing costs and limitations of  Web interaction. On the other hand, we are exploring possibilities to host validation workflows that would allow at least a sanity check of the user-supplied mesh, before its adoption for simulation.
VERCE has initiated a useful catalogue of meshes and corresponding tomographical models, which others may use and to which they may 
contribute. Such Web accessible resources 
will be key elements in motivating collaboration, while a mesh generating workflow is being investigated.  

The event parameters are obtained from sources such as the GCMT or the INGV catalogues, which use different estimation methods, or are the result of local analyses. 
These will increase in complexity as the models of the physics of the slip processes are developed.
Geodesy, interferometry (InSAR---Interferometric synthetic aperture radar) to measure ground motion and space-image data may be used to support further analyses in those cases.
To orient the simulation setup and results, \eg\ as visualisations or videos, we make use of more geospatial and geopolitical context-setting data and appropriate visualisation services, typically 
those supporting OGC\footnote{\url{www.opengeospatial.org}} standards as mandated by the INSPIRE directive \cite{INSPIRE2007}. 
These services are provided for public use from a number of organisations and we currently integrate those offered by European initiatives including OneGeology\footnote{\url{www.onegeology.org}} 
and SHARE\footnote{\url{www.share-eu.org}} and governmental institutions such as the KNMI, providing respectively geological information, regional hazard maps and 
other environmental visualisation overlays.
Other disciplines, \eg\ civil engineering, mineral extraction and urban planning, may access the resulting data, for instance to investigate anthropogenic seismicity. 
They may even use the platform for their regional studies, adopting private models with restricted access policies. 
This last  case is supported by the authorisation-based access control to the data, which is in place throughout the storage elements of the platform.

\shorten\shorten
\subsection{Provenance-driven tools} \label{ProvenanceTools.ssec}
\shorten\shorten
\hidempaNote{How provenance collection is organised and controlled. How provenance is integrated. How tools navigate provenance 
graphs and launch actions on encountered data.}

To achieve quality, science must be replicable, \ie\ the same or a different scientist should be able to repeat a digital experiment.
This is enabled in VERCE by consistently handling provenance records across the diversity of subsystems. This has a further 
payoff; it results in a consistent model of the relationship between methods and data that is presented and exploited by tools 
that help scientists explore and validate their data and processes.

The VERCE science gateway combines workflows interactively behind its tailored user interfaces. 
This integration of specific and generic tools led to a monitoring and validation system
that is used to evaluate \emph{at run-time} the progress of the computations for all of the interactive services accessed via the gateway. 
This needs to present information to users specific to their application domain that enables them to make decisions 
that have immediate effects. 
For instance, if we consider long-running cross-correlation analyses or high-resolution simulations, 
detecting errors in results or metadata  during the run, can trigger reactions that can save users from 
unfruitful waits and from wasting energy or expensive computing cycles. Such direct control makes their research 
more efficient and progress faster. 

Pauw \etal\ \cite{Pauw2010} argue that the generation of huge quantities of provenance traces is excessively demanding for an effective visualisation tool,
and it overloads users. 
To reduce this cognitive overload and to provide orientation when investigating properties of the processes and  data, we 
specify and implement a Web API, targeting common interrogation patterns, which are 
combined and orchestrated via a user interface.  
This provides unified access to fine-grained provenance, as well as summary information and both user and domain-specific metadata, 
as recommended in PROV-DM~\cite{prov-primer}.  
Its  implementation takes advantage of the flexibility of NoSQL technologies, such as MongoDB.

\begin{figure*}
\begin{center}
\shorten\shorten
\includegraphics[width=16cm]{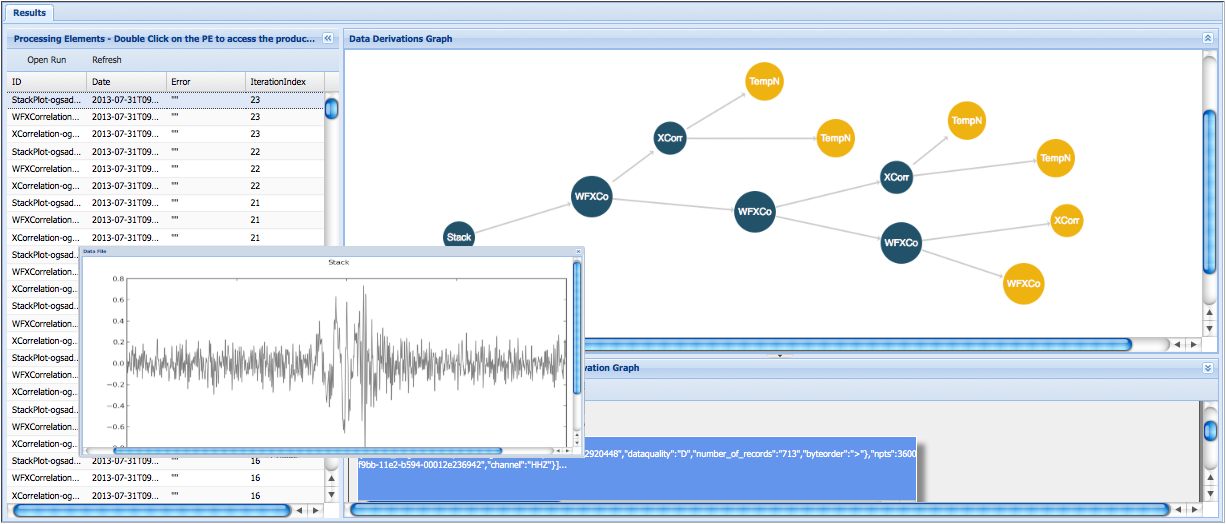}
\shorten
\caption{Provenance Explorer GUI: Visualisation of the interactive derivation graph and data preview for an intermediate trace of a cross-correlation workflow. 
The dots represent data at a transformation stage. Yellow dots are expandable to allow the further navigation through the dependency graph.\shorten\shorten}
 \label{fig:fw-derivgraph}
\end{center}
\end{figure*}

Users search for runs and for data products by submitting metadata queries as value ranges. 
They examine search results by browsing the metadata associated with the data stream, including the data relationships.
This is illustrated in Fig.~\ref{fig:fw-derivgraph}. 
The dynamics of this exploration develops very detailed insight into a workflow's logic. 
Generated files are linked by the provenance metadata and made available for visualisation or download, by accessing the underlying data management system. 
For rapid and large size download operations, a bulk download script matching the search results is automatically generated. 
This allows users to fetch their selected data in batch mode, into their local facilities adopting  appropriate data transfer protocols. 
Portability is supported by enabling users to export full traces of runs in W3C-PROV format. 
This will foster the hand over of a final research product to institutional and multidisciplinary archives, 
where provenance is included in their data-curation processes.

Provenance systems typically collect metadata about the data transformations~\cite{Foster:2003:VDG:1116842.1116845} during workflow execution. 
We have introduced additional features to trigger actions using third-party components to react to provenance information. 
An operational example in the VERCE, is the runtime shipment of the intermediate data from a DCI to external systems, 
to facilitate rapid diagnostic evaluation or visualisation. 
In many cases this behaviour can't be instrumented within the workflow itself because of the lack of external connectivity from HPC clusters.


 \shorten\shorten\shorten
 \section{Building the VERCE VRE} \label{experiencesBuildingTheVRE.sec}
 \shorten\shorten
 \hidempaNote{A selection of experiences encountered setting up and running the VERCE platform. Pick a few, some illustrate 
 problems surmounted and some that have so far proved intractable. In these cases the work around we use if we have 
 found one. We promise to ``reports on the difficulties encountered setting 
up this coherent, European-wide accessible research platform---those overcome and those persisting."}

Science gateways are one of the most discussed, often criticised, but yet required components of a VRE. 
The most challenging task that we had to face was how can we make it actually useful and transparent for our community. 
Our solution was to engage people in \emph{task forces} where committed users, developers and service providers 
use co-design and co-development to overcome limitations that seismologists would otherwise experience. 
This changed the focus from HPC production environments and community data services, 
where middleware standards, security procedures and connectivity dominated, to a user-focussed scenario.  
Without these task forces, users would have run away rather than engaged with the VERCE platform.

An example task force focused on misfit calculation. It involved two computational seismologists, an 
expert developer of the \pw{ObsPy} Python library, two \dfp\ experts, a data expert, an expert in provenance and the 
developer of interaction templates and portlets. After intensive preparation, it had a three-day kick-off 
meeting, at which current systems were tried and ideas for improvements were hatched and critical 
steps evaluated. Two months of intense development and collaboration followed, with daily 
communication between pairs attacking subgoals. 
A further face-to-face accelerated the integration of the components, towards exhaustive tests and refinement before production deployment.
It is now an integrated element of the VERCE science gateway.
These task forces adopt agile development methods but have a greater focus on teasing out what new scientific methods 
are becoming feasible as the digital context improves.

We had to accommodate users with different levels of expertise. 
Some require an advanced programming environment.  
Others require results from simulations with a variety of different input configurations and parameterisations. 
This requires significant house keeping, detailed technical knowledge and awareness of the potential offered by the external 
domain-specific services and facilities which are out there to serve the community with data and tools. 
During our training sessions we discovered that, regardless of their level of expertise, users are 
often unaware of or ignore such services. By automatically organising data movement, house keeping and 
resource use, VERCE not only accelerated and simplified the functional steps, it also migrated work to suitable targets. 
This was demonstrated from the trainees questions which, as a result, were focused on the parametrisation of the simulation and the modelling methodologies. 
Also the  novel interactive paradigms for result validation via metadata and provenance exploration, were used enthusiastically. 

Some parts of the VERCE e-Infrastructure implementation are shown in \tref{VERCEeInf}\footnote{All subsystems in that table 
are operational in the actively used system. Three are still under development: (i) misfit analysis still requires some user interfaces 
in the science gateway and is used directly at present, (ii) the pre-configured library of pipelines continues to have additions, and (iii) 
the bulk data ingest kit is used directly and not yet packaged in the science gateway. Security requirements remain an issue as described below.}.
The first two columns use technology, such as simulation codes and FDSN Web services, that is specific to computational and data-intensive seismology respectively.
We anticipate that placing equivalent data, technology and functions here for another discipline would now be straightforward as the relationships with 
the other two columns have been established. In many cases there are templates and workflows that could be suitably modified and reused.
These last two columns provide an e-Infrastructure and many generic services that would be equally necessary and applicable in other contexts.
As far as possible existing technologies, \eg\ \pw{SPECFEM3D}, \pw{Liferay}, \pw{DCI-Bridge} and \pw{iRODS},  and services have been used to reduce delays and share maintenance costs.
This should be taken further as systems such as the products of the EUDAT projects become available.
In most cases these need to be deployed on hardware, systems software and networks that are shared or used for multiple purposes.

Resilience and sustainability depend on appropriate insulation from the changes in software services, data representations, network protocols and hardware architectures.
We start with traditional insulation, such as:
\begin{inparaenum}[\itshape a\upshape)]
\item	high-level languages, \eg\ Python, and operating systems, \eg\ Linux, to span diverse computers, 
\item computational frameworks, \eg\ UNICORE and Globus, to span operating environments,
\item data management frameworks, \eg\ iRODS, to span storage systems,  
\item job submission and workflow enactment frameworks, \eg\ DCI-Bridge, protecting against diversity and change in computing services, and
\item Web and document management, \eg\ Liferay, providing a consistent environment for interaction development.
\end{inparaenum}
VERCE builds on this to protect investment and composes large algorithmic elements using \pw{WS-PGRADE/gUSE} workflow enactment system, 
that works with all the DCIs that we encountered.
It also handles many of the detailed requirements of credential, authority and identity management.
VERCE composes smaller-scale tasks and fine-grained operations using \dfp\ because of its low overheads and multiple mappings 
to multiple data-intensive middleware systems.
These two composition methods offer two advantages:
\begin{inparaenum}[\itshape a\upshape)]
\item		the development of templates that can be reused for other domains and adapted at a high level for new target environments, and
\item 	connection with larger user communities that will invest in mappings to new technologies.
\end{inparaenum}
Inevitably, using a series of insulating layers also reduces the scope for innovation and new approaches as identified by Gesing \etal\ \cite{Gesing:2014:WDN:2691175.2691188}.

Management of security, identity, authorisation and accounting is still problematic. 
We attempt to provide a consistent continuum of facilities and resources so that 
researchers can experiment and innovate freely in a low-cost, low risk environment and move their work fluently into powerful production contexts. 
The former context imposes very few security hoops for researchers to jump through.
The latter require careful protection because of their expense and potential for misuse.
In VERCE we try to combine open-data and free to use services, with security mechanisms adopting the VOMS\footnote{\url{repository.egi.eu/2012/07/10/voms-2-0-8/}} model, 
which is an advanced and effective system, technically and conceptually, based on the grid X.509 certificates. 
It presents some weakness, \eg\ in the definition of roles so that they are semantically consistent across contexts. 
It is adopted by service providers involved in the EGI initiative, but not yet fully recognised by others. 
Consequently, today there are different security hoops to leap through in almost every context, and many of them require lengthy set up. 
A wide variety of computational and data-intensive researchers and innovators face this problem.
It is surprising that there is not a European-wide or internationally accepted, sufficiently trusted, sufficiently multi-scale robust solution to achieve easy migration between research contexts.
The impediments are largely organisational and political.
Lack of it continues to inhibit uptake; thus depriving researchers of significant potential.
Steps to resolve this disharmony need to involve many communities, many categories of provider and all
e-Infrastructure campaigns; VERCE exposes the issue, others must solve it.

\begin{table*}
\shorten\shorten\shorten
\caption{Sub-systems implementing the VERCE e-Infrastructure}\label{VERCEeInf.tab}
\begin{tabular}[h]{|p{4cm} |p{4cm} |p{4cm} |p{4cm} |}
\hline
\multicolumn{1}{|c|}{Computational seismology} & \multicolumn{1}{c|}{Data-intensive seismology} & \multicolumn{1}{c|}{Data management} & \multicolumn{1}{c|}{Framework} \\
\hline
\hline
\pw{SPECFEM3D} wave simulation & \pw{ObsPy} Python library 	& \pw{iRODS} federated store for seismic data 	& Science gateway using \pw{Liferay} and \pw{gUSE} \\
\hline
Misfit analysis library				& Seismology PE libraries 	& \dfp\ Python system 		& Job submission via \pw{DCI-Bridge} \\
\hline
Pre- and post-processing pipelines & Seismic trace processing pipelines & Linking seismic data and metadata & Gateway deployed on clouds \\
\hline
Mesh and model uploads and catalogues	& Pipeline editing and creation	& \dfp\ on local machines, HPC and specialised clusters and cloud & Computations on HPC and institutional clusters \\
\hline
Integration of moment tensor sources & Production and capture of metadata & Provenance handling and exploitation & Security requirements compliance \\
\hline
Production and capture of metadata & Data ingest tools  & \pw{iRODS} and \pw{MongoDB} for metadata & Scientific workflows \\
\hline
		&		& GridFTP for bulk data transport  & Provenance management \\
\hline
\end{tabular}
\shorten\shorten
\end{table*}

 \shorten\shorten
 \section{Benefits from the VERCE VRE} \label{VREbenefits.sec}
 \shorten\shorten
 \hidempaNote{For each of the categories of use of the VERCE VRE identify successes in the innovative phases and 
 production phases of research. Show acceleration of automated methods, improved agility and growing flexibility.}
 
 Researchers report completing computational seismology tasks in a few hours that would have previously taken days or weeks.
 They use a wide array of computational
 facilities to satisfy their simulation
                      requirements. Daily research activities are		
                      satisfied on the institutional resources where
                      they have routine access while ground breaking
                      research activities that require higher
                      performance capabilities are achieved on national
                      and European facilities, such as PRACE. The VERCE
                      framework enables the researchers to work
                      seamlessly on such facilities without having to
                      learn the heterogeneity of the underlying
                      environment. With VERCE's support, potentially any
                      associated code sets, workflows and data can be
                      administered. 

Indeed, one leading seismologist described the progress in this way, ``{\em We have experienced three phases of computational seismology.
\begin{inparaenum}
\item		Write your own code, parallelise it and do science organising everything yourself. It doesn't work anymore---it takes too long and the hardware is too complicated.
\item		Download community code developed by a team. Install it and run your jobs. It still requires organising everything yourself and much technical knowledge.
\item		Use a platform like VERCE.
\end{inparaenum}
The big benefits from the platform are you can focus on your science and productivity improves by one or two orders of magnitude.}" \cite{Igel2015}
 
The VERCE platform tackles commonly experienced cross-disciplinary issues encountered by users as they develop, test
and run {\it in silico} experiments and simulations. 
It provides immediate access to tools to explore the metadata, the dependencies and the diagnostic logs related to any application 
activated via the science gateway. 
We move from a very practical and science-driven way of setting up an {\it in silico} experiment to a more flexible and 
comprehensive way of supporting users during the data preparation and result evaluation, 
where data and methodologies are merged into a single exploratory space. 

Context specific helpers improve the uptake of such tools. 
For instance, the current search interface over the experiments, adopting metadata relevant to the community's context, would in future make use of a tag cloud of terms.  
Their relevance can be weighted according to the user's scope, without hiding foreign metadata vocabularies which might have been used to describe results 
obtained by different applications, using the same workflow components. 
This contextual information can be extrapolated by the analysis of provenance data, particularly with the dynamic participation of users, 
which should enable consistent and effective acquisition of the information which they seek. 

These adaptive and general  interfaces, aimed at exploring the computational processes, make it easier for researchers to exploit their results. 
For instance, combining exploratory actions with data-shipment  to other facilities, exposes them to a much bigger space of data services; 
still within the framework of a federated authentication infrastructure. 
This is achieved by the automated generation of helpers that handle security and transfer protocols. 
Such provenance-empowered `awareness tools', provide users with the opportunity to learn about the computational process behind the tool they use; 
even when processes fail they yield rapid feedback. This does not detract from their primary focus and the principle uses of results. 
 
 \shorten\shorten
 \section{Conclusions and future directions} \label{finale.sec}
 \shorten\shorten
 \hidempaNote{Assess the overall achievements absolutely and in comparison with contemporary work.
 Identify areas where more research is needed, others where more cooperation is the answer, and a few where 
 it is apparently only necessary to pursue a reasonably well identified development plan.}
 
 A key and sustained asset of VERCE has been very close collaboration between geoscientists and computer scientists.
 This was a crucial requirement for success; without it the task forces would not have worked. 
 But it is also a key long-term goal, VERCE has to deliver a VRE where a growing number of seismologists and solid-Earth scientists will continue to advance their 
 science and increase the power of their platform for doing their science. This has required that the payoff to seismologists from students to professors had to be manifest,
 but there also had to be benefits for the teams of ICT experts who run the dedicated and shared resources seismologists need.
 Their payoff, is systems that interwork without their intervention, and templates, workflows and tools that automate many of their recurrent tasks.
 VERCE has achieved this positive feedback loop and demonstrated it for its computational and data-intensive  communities.
 
 This has been achieved by making progress on all four research tracks identified by Mattmann as critically important for future data science, based on many years of experience at NASA and at the 
 Apache Software Foundation \cite{Mattmann2014}.
 \shorten
\begin{itemize*}
\item{\em Rapid scientific algorithm integration:} The platform can easily support new simulation codes, thanks to a flexible design of configuration schemas and interactive components, 
	which evolve around interoperable formats and functional abstractions. In that respect, the next target will be 
	\pw{SPECFEM3D\_GLOBE}\footnote{\url{geodynamics.org/cig/software/specfem3d_globe/}} as an alternative simulation code to improve support for global-scale waveform propagation studies. 
	In the data-intensive contexts, new processing algorithms can be  quickly put in practice, thanks to the portability of \dfp{} which allows for their flexible evaluation across environments. 
	We will improve the integration of the platform with registry and software packaging mechanisms, to better support the interactive creation of data-intensive workflows, 
	which can then be deployed and executed under the full control of users, including provenance recording and metadata customisation.

\item{\em Intelligent data movement:} This is key to meeting operational constraints and to achieving performance. 
	We automatically handle getting the data to the right place at the right time, releasing resources in a timely manner 
	while automatically collecting metadata and provenance records, so that users can find and navigate their data. 
	Performance includes user requirements for immediate feedback and control of an experiment. 
	We can instruct the platform to produce and selectively make available intermediate results at runtime, 
	fostering the rapid reaction of humans, as well as software agents external to the DCI. 
	This is possible as we address one of the most important limitations Marrmann identified---the lack of effective metadata management. 
	We do not considered HPC jobs as closed systems, but rather active entities which can trigger behaviours and state changes in external services while they run. 
	This results from combining runtime provenance analysis with selective data movements. 
	We will develop more dynamism and semantics into this framework, in order to exploit diverse and reactive agents. 
	
\item{\em Use of Cloud Computing:} Clouds are used in VERCE for the operation and sustainability of the platform itself, 
	rather then for providing elastic resources and software stacks typical in \emph{Cloud Marketplaces} that provide virtual appliances. 
	We use VM deployments where we can, with mappings onto advanced data-intensive technologies. 
	This provides the system administration gains even where resource allocation takes different forms.
	Many HPC providers are evaluating the impact of virtualisation on their architectures. 
	A key achievement delivered is to recognise both the potential and drawbacks of different virtualisation models 
	and to allow research activities to move fluently between them.
	
\item{\em Harnessing the power of open source in software development for science:} VERCE itself is a demanding user of open source projects, 
	such as \pw{gUSE}, \pw{iRODS}, \pw{Liferay}. 
	Several of its developments are becoming open source projects as a policy for the sustainability of its subsystems, 
	see \ref{experiencesBuildingTheVRE.sec}. 
	\eg\ the data-intensive framework \dfp{},  will be packaged to reach a wider 
	audience of users and developers. 
\end{itemize*}
 
These technical directions must be aligned with organisational
strategies. The VERCE community will continue to
grow and investment in training for established and new seismologists to 
alert them to its potential will need commensurate growth.
Seismology continuously strives to refine the resolution
of local, regional and global Earth models, as current resolutions are very often one-or-two
orders of magnitude short of imaging the fundamental geophysical processes.
Taking advantage of rapid advances in geodesy and geodynamic modelling
is another challenge. The breadth of solid-Earth sciences
will be extended with an increase in the diversity of data,
simulations, and interconnected resources to enable this, in
part in the context of EPOS. To achieve such extensions
will require resources, but far less than starting again in
each domain and merging later. There is a challenge to raise
the {\em activation energy} to make that transition, which requires
clarity about the benefits. This paper offers a glimpse of
much future potential.
 
{\bf  Lessons learned}:
 \begin{enumerate*}
 \item There are many resources: computational, data and working methods formalised as code and workflows, that have 
 	many technical and organisational variations that will continue because of education, investment and commitments.
	Providing a means for researchers to use these in combination without encountering the underlying complexity 
	opens up new research avenues and improves their productivity dramatically.
	It also increases the opportunity for experts in different subdomains to pool their insights and skills.
 \item It is difficult to find organisational and technical paths to reach that goal. Establishing short-lived, agile multi-disciplinary task forces
 	to address specific issues is key to developing the depth of understanding and commitment to make progress.
 \item The orders of magnitude gain in productivity have to be balanced with flexibility to enable
 	researchers to create, refine and drive whatever innovative approach they choose.
	The challenges in each domain are so great that flexibility must be retained and exploited through multiple methods.
\item The independent paths to improvement of each contributing organisation, sub-discipline and technology
	are necessary for sustained vitality. Therefore, their composition must depend on mappings that are
	easily revised.
\item {\em Jam tomorrow} is not enough. Vision and strategic thinking are needed to achieve convergence, but 
	immediate paths from current contexts to working examples with evident payoffs are necessary to maintain 
	engagement, avoid unacceptable disruption and refine the target vision.
 \end{enumerate*}
 
 {\bf Open issues}:
 \begin{enumerate}
 \item It is not possible to completely hide different usage policies such as identification, authentication, authorisation and accounting, 
 	because light-weight mechanisms to help large numbers of students and citizen scientists don't offer the trustworthy protection 
	needed for major facilities used by moderate numbers of specialists. This balance between convenience and verified security 
	requires treatment for many application domains as they embrace multi-disciplinary, multi-orgainisational and multi-national 
	collaboration. However, this fundamental issue is exacerbated by a legacy of diverse security mechanisms in use. The full breadth
	of institutions involved must at least circumvent these legacy impediments to collaboration. The critical requirements are the development of 
	mutual trust across all of the organisations involved and investment in replacing archaic mechanisms with easily used
	and easily managed interworking mechanisms.
 \item The sustained delivery of e-Infrastructures is needed to warrant researchers becoming dependent on them. This will involve 
 	the maintenance of many software elements, some of which are introduced above. Just to keep them running with no enhancements, 
	software has to be revised to meet changes in its operational context. Normally there are also pressures to address bugs and 
	to introduce new features. Most components should be widely used and open source, so that many communities can 
	share the necessary maintenance. As far as possible, the e-Infrastructure should use such components and replace them 
	if they do not have a thriving community. This leaves three forms of software maintenance that every e-Infrastructure community 
	needs to be responsible for:
	\begin{inparaenum}[\itshape a\upshape)]
	\item their fair contribution to the multi-community software elements,
	\item the mappings to and integration between the common software elements to meet their specific needs, and
	\item on hopefully the rare occasions when a major element needs to be replaced by a thriving alternative, the 
		integration of that alternative.
	\end{inparaenum}
	Today this maintenance investment is only available for novelty items and recognised simulation codes. 
	Many other software elements need maintenance 
	for the investment in e-Infrastructure to survive and for the improved research environment to be sustained.
	Funders, organisations providing facilities, e-Infrastructure builders and VRE developers need to form alliances
	to achieve this for the research infrastructures that are strategically important.
\item The tools and methods for setting up and maintaining flexible mappings for e-Infrastructures such as VERCE, 
	where many organisations independently manage and evolve
	their own services, need to be improved substantially.
	Today the inclusion of new facilities and the setting up of mappings to exploit those resources requires too much effort and expertise.
	The abstractions do not sufficiently isolate the mappings and optimisations for production 
	require steering by specialists. 
	We anticipate that ESFRI investments, such as EPOS, will meet these issues on a
	very significant scale. 
	
	We anticipate progress on the following fronts:
	\begin{enumerate}
	\item	Improved abstract notations for describing computational and data-intensive scientific methods
		with good tools for their creation, refinement and composition. These tools will map down to
		multiple underlying workflow systems and middleware e-Infra\-structures. The notation will 
		represent clearly the logical steps from a domain researcher's viewpoint and identify key factors, such as the identity
		of data sources or input parameters, so that mappings have sufficient information to automatically optimise and adapt. However,
		target and operational details will be excluded, so that once developed the methods are long-lived
		and highly portable. The notations will be easily understood by domain scientists so that they can
		directly create and improve their own methods. Composition will support combinations of sub-methods
		developed by different experts. Computational and data-science specialists will provide
		and optimise common sub-methods and method patterns.
	\item	Much improved mechanisms for describing target computational, storage, network and middleware
		systems that are precise and sufficient to form the basis for automated federation across
		evolving subdomains. The investment in these descriptions will initially be complex, as it was in
		the Grid days, but ultimately there should be a substantial body of described systems, and good
		tools for creating and validating those descriptions. The current standards for describing digital
		devices indicate that such a scale and complexity can be tackled. As subsystems are
		introduced or replaced the set of descriptions of potential targets will grow. As properties of services,
		subsystems and middleware change, their descriptions will change in such a way that the
		mappings and inter-working relationships are automatically maintained in nearly every case.
		Once adequate descriptions are in place research will be needed to develop algorithms that
		adapt mappings while sustaining method semantics.
	\item Developing dynamic mapping algorithms that optimally transform the abstract scientific methods onto the
		contributing federation of resources, taking account of user rights and entitlements, of previous
		performance data, of operational load and of data locality. The cost models will be well adapted to
		minimise the environmental impact of the workloads and to yield responsive interactive working.
		Operational data will facilitate assessment of the validity of resource allocation, e.g. whether a
		scientific result warrants the environmental impact.
	\end{enumerate}
	
	This will be achieved through international open-source projects, business-led R\&D and a cluster of
	research projects combining fundamental and engineering advances. It is a long-term goal, which all
	research communities and e-Infrastructure builders should keep in mind. This investment is necessary to
	sustain research that depends on collaborations between independent resource-providing organisations,
	i.e. those supporting most of today's interdisciplinary research. These organisations deliver federations
	of diverse DCIs. The new approach is required to sustain these federations in the context of evolving
	technology and changing operational practice. Without such advances it will be necessary to commit
	unaffordable investment in highly skilled ICT specialists in order to continue to support researchers
	addressing society's strategic and urgent challenges using their integrated virtual research environments.
 \end{enumerate}


\shorten
\section*{Acknowledgments and dedication}
 VERCE is supported by the the EU project RI 283543.
 The 21 authors are those who contributed directly to this paper.
 They thank the many others in VERCE who contributed to the ideas, to the software, to setting up and running 
 the services, and to supporting users. They also thank the many who have developed seismic services and 
 simulation codes over the last two decades, and the students and users beyond VERCE who have contributed 
 their feedback. They dedicate this paper to the memory of Torild van Eck, whose premature death robbed him 
 of the chance to share VERCE's success and to take a leading role in this paper. His sustained commitment 
 to collaboration, his vision and quiet diplomacy, as leader of ORFEUS\footnote{\url{www.orfeus-eu.org}} at KNMI, built the inter-disciplinary 
 cooperative spirit that led to our success.
  

\bibliographystyle{IEEEtran}

{\scriptsize{		
\setlength{\parskip}{0.0pt plus 0.3ex}		
\setlength{\itemsep}{0.0pt plus 0.3ex}		

\bibliography{wf-survey}

}
}



\end{document}